\documentclass{article}
\usepackage{amsmath,amssymb,color,url}
\usepackage{amsthm}
\allowdisplaybreaks
\newtheorem{thm}{Theorem}[section]
\newtheorem{cor}[thm]{Corollary}

\newtheorem{prop}[thm]{Proposition}
\newtheorem{defn}[thm]{Definition}
\newtheorem{rem}[thm]{Remark}
\newtheorem{ass}[thm]{Assumption}
\newtheorem{ex}[thm]{Example}
\def\fin   {\hfill{$\Box$}\vspace{5mm}}
\def\l     {\left}
\def\r     {\right}
\def\la    {\langle}
\def\ra    {\rangle}

\def\sgn   {\mathop{\mathrm{sgn}}\nolimits}

\def\calF  {{\cal F}}

\def\bbC   {{\mathbb C}}
\def\bbD   {{\mathbb D}}
\def\bbE   {{\mathbb E}}

\def\bbP   {{\mathbb P}}
\def\bbR   {{\mathbb R}}
\def\ve    {\varepsilon}
\def\vp    {\varphi}
\def\vt    {\vartheta}
\def\tP    {{\bbP^\ast}}
\def\tN    {{\widetilde{N}}}
\def\olh   {{\overline{h}}}
\def\olt   {{\overline{t}}}
\def\olx   {{\overline{x}}}
\def\olxi  {{\overline{\xi}}}
\def\whC   {\widehat{C}}
\def\whH   {\widehat{H}}
\def\whM   {\widehat{M}}
\def\whS   {\widehat{S}}
\def\whV   {\widehat{V}}
\def\whg   {{\widehat{g}}}
\def\whmu  {{\widehat{\mu}}}

\def\bfPsi {{\mathbf \Psi}}

\begin{document}
\title{A Clark-Ocone type formula via It\^o calculus and its application to finance}
\author{Takuji Arai\footnote{Department of Economics, Keio University, 2-15-45 Mita, Minato-ku, Tokyo, 108-8345, Japan (arai@econ.keio.ac.jp)} and
        Ryoichi Suzuki\footnote{Department of Mathematics, Keio University, 3-14-1 Hiyoshi, Kohoku-ku, Yokohama, 223-8522, Japan (r\_suzuki@z3.keio.jp)}}
\maketitle

\begin{abstract}
An explicit martingale representation for random variables described as a functional of a L\'evy process will be given.
The Clark-Ocone theorem shows that integrands appeared in a martingale representation are given by conditional expectations of Malliavin derivatives.
Our goal is to extend it to random variables which are not Malliavin differentiable.
To this end, we make use of It\^o's formula, instead of Malliavin calculus.
As an application to mathematical finance, we shall give an explicit representation of locally risk-minimizing strategy of digital options for exponential L\'evy models.
Since the payoff of digital options is described by an indicator function, we also discuss the Malliavin differentiability of indicator functions with respect to L\'evy processes.
\end{abstract}


%
%
\setcounter{equation}{0}
\section{Introduction}
An explicit martingale representation for random variables described as a functional of a L\'evy process will be given by using It\^o's formula, instead of Malliavin calculus.
As an application to mathematical finance, we provide a representation of locally risk-minimizing (LRM) strategy of digital options for exponential L\'evy models.

Consider a square integrable $1$-dimensional L\'evy process $X$ expressed as
\begin{equation}\label{eq-X}
X_t=X_0+\mu t+\sigma W_t+\int_{\bbR_0}x\tN([0,t],dx)
\end{equation}
for $t\geq0$, where $X_0\in\bbR$, $\mu\in\bbR$, $\sigma\geq 0$ and $\bbR_0:=\bbR\setminus\{0\}$.
Here, $W$ is a $1$-dimensional standard Brownian motion, $N$ is a Poisson random measure; and $\tN$ is the compensated measure of $N$, that is, it is represented as
\[
\tN(dt,dx)=N(dt,dx)-\nu(dx)dt,
\]
where $\nu$ is the L\'evy measure of $N$ satisfying $\int_{\bbR_0}x^2\nu(dx)<\infty$.
For a time horizon $T>0$ and a measurable function $f:\bbR\to\bbR$ such that $f(X_T)$ is square integrable, the martingale representation theorem implies that
\begin{equation}\label{eq-CO}
f(X_T)=\bbE[f(X_T)]+\int_0^Tu^f_sdW_s+\int_0^T\int_{\bbR_0}\vt^f_{s,x}\tN(ds,dx)
\end{equation}
for some predictable processes $u^f$ and $\vt^f$.
The Clark-Ocone theorem (see, e.g., Theorem 3.5.2 of Delong \cite{Delong}) says that $u^f$ and $\vt^f$ are described as conditional expectations of Malliavin derivatives of $f(X_T)$
if $f(X_T)$ is Malliavin differentiable, that is, $f(X_T)$ belongs to the space $\bbD^{1,2}$ defined in Section 2.2 of \cite{Delong}.
On the other hand, when $f(X_T)$ is not Malliavin differentiable, e.g., ${\bf 1}_{\{X_T\geq 0\}}$ with $\sigma>0$, there is no way to calculate $u^f$ and $\vt^f$ explicitly.
In this paper, we aim to give concrete representations of $u^f$ and $\vt^f$ by using It\^o's formula, instead of Malliavin calculus,
under some conditions which have nothing to do with the Malliavin differentiability of $f(X_T)$.
To this end, regarding the conditional expectation $\bbE[f(X_T)|X_t=x]$ as a function on $(t,x)\in[0,T]\times\bbR$, denoted by $F$, we apply It\^o's formula to $F$.
As a result, we obtain a Clark-Ocone type formula \eqref{eq-CO} in which $u^f$ and $\vt^f$ are given as a partial derivative and a difference of $F$, respectively.

Using the obtained Clark-Ocone type formula, we shall provide a representation of LRM strategy of digital options for exponential L\'evy models in the second part of this paper.
Remark that LRM strategy is a well-known quadratic hedging method, which has been studied very well for about three decades, for contingent claims in incomplete markets.
Consider a financial market composed of one risk-free asset with interest rate $r\geq0$ and one risky asset
whose fluctuation is described by the following exponential L\'evy process $S$:
\begin{equation}\label{eq-exp}
S_t:=e^{rt+X_t}
\end{equation}
for $t\geq0$.
Then, the payoff of digital options is expressed as ${\bf 1}_{\{S_T\geq K\}}$ with $K>0$.
Note that we need to assume some conditions on $X$ in order to use our Clark-Ocone type formula.
Considering three L\'evy processes: Merton jump diffusion, variance gamma (VG) and normal inverse Gaussian (NIG) processes,
as examples of representative L\'evy processes frequently appeared in mathematical finance, Merton jump diffusion and NIG processes satisfy our conditions, but VG processes do not.
However, it is known that ${\bf 1}_{\{X_T\geq c\}}\in\bbD^{1,2}$ for $c\in\bbR$ if $\int_{\bbR_0}|x|\nu(dx)<\infty$ and $\sigma=0$ such as VG processes.
Thus, when $X$ is a VG process, a representation of LRM strategy of digital options are given from Example 3.9 of Arai and Suzuki \cite{AS},
which has provided a general expression of LRM strategies for exponential L\'evy models by means of Malliavin calculus.
On the other hand, as is well-known, ${\bf 1}_{\{X_T\geq c\}}\notin\bbD^{1,2}$ whenever $\sigma>0$ such as Merton jump diffusion processes.
Moreover, we shall show in the last part of this paper that ${\bf 1}_{\{X_T\geq c\}}\notin\bbD^{1,2}$ holds if $\int_{\bbR_0}|x|\nu(dx)=\infty$ and $\sigma=0$ such as NIG processes.
In summary, our result in the second part provides the only way to calculate LRM strategy of digital options for the case where $X$ is a Merton jump diffusion process or an NIG process.

The remainder of this paper is organized as follows:
A Clark-Ocone type formula for $f(X_T)$ is shown in Section 2.
In Section 2.4, explicit martingale representations for various functions $f$ will be introduced.
Section 3 is devoted to LRM strategy of digital options.
In the last subsection, we discuss the Malliavin differentiability of indicator functions with respect to L\'evy processes.

%
%
\setcounter{equation}{0}
\section{Clark-Ocone type formula}
For a L\'evy process $X$ described by \eqref{eq-X} and a measurable function $f:\bbR\to\bbR$, we aim at providing a Clark-Ocone type formula for $f(X_T)$ using It\^o's formula.

\subsection{Preparations}
Before stating our main theorem, we need some preparations.
Denoting the characteristic function of $X_T-X_t$ by $\phi(t,z)$ for $(t,z)\in [0,T]\times\bbC$, we have and denote
\begin{align}\label{eq-psi}
\phi(t,z) &:= \bbE[e^{iz(X_T-X_t)}] = \bbE[e^{iz(X_{T-t}-X_0)}] \nonumber \\
          &=  \bbE\l[\exp\l\{iz\l(\mu(T-t)+\sigma W_{T-t}+\int_{\bbR_0}x\tN([0,T-t],dx)\r)\r\}\r] \nonumber \\
          &=  \exp\l\{(T-t)\l(iz\mu-\frac{\sigma^2z^2}{2}+\int_{\bbR_0}(e^{izx}-1-izx)\nu(dx)\r)\r\} \nonumber \\
          &=: \exp\l\{(T-t)\psi(z)\r\}
\end{align}
by the L\'evy-Khintchine formula (see, e.g., (8.8) in Sato \cite{Sato}).
Now, we give assumptions on $X$ as follows:

\begin{ass}\label{ass1}
\begin{enumerate}\renewcommand{\labelenumi}{(\arabic{enumi})}
\item There exists $\alpha>0$ such that $\bbE\l[e^{\alpha X_T}\r]<\infty$.
\item For any $\alpha>0$ with $\bbE\l[e^{\alpha X_T}\r]<\infty$ and any $t\in[0,T)$, there exists an integrable function $h_t(v)$ on $\bbR$ such that
      \[
      |\phi(\olt,iz_v)|\l(1+|z_v|+\frac{1}{|z_v|}\l|\int_{\bbR_0}(e^{-z_vx}-1+z_vx)\nu(dx)\r|\r)\leq h_t(v)
      \]
      for $\olt\in[\frac{t}{2},\frac{T+t}{2}]$, where $z_v=iv-\alpha$.
\end{enumerate}
\end{ass}

\begin{rem}
By Proposition 3.14 of Cont and Tankov \cite{CT}, the above condition (1) is equivalent to the following two conditions:
\begin{itemize}
\item[(1)${}^\prime$] There exists $\alpha>0$ such that $\bbE\l[e^{\alpha X_t}\r]<\infty$ for any $t\in[0,T]$,
\item[(1)${}^{\prime\prime}$] There exists $\alpha>0$ such that $\displaystyle{\int_{\{|x|\geq1\}}e^{\alpha x}\nu(dx)<\infty}$.
\end{itemize}
On the other hand, under (2), $X_T$ has a bounded continuous density from Proposition 2.5(xii) of \cite{Sato}.
\end{rem}

We introduce three examples of L\'evy processes, which are frequently appeared in literature for mathematical finance, and discuss whether or not they satisfy Assumption \ref{ass1}.

\begin{ex}[Merton jump diffusion processes]\label{ex-Merton}
A L\'evy process $X$ described by \eqref{eq-X} is called a Merton jump diffusion process, if $\sigma>0$ and
\[
\nu(dx)=\frac{\gamma}{\sqrt{2\pi}\delta}\exp\l\{-\frac{(x-m)^2}{2\delta^2}\r\}dx,
\]
where $m\in\bbR$, $\delta>0$ and $\gamma>0$.
In this case, $X$ consists of a Brownian component and compound Poisson jumps with intensity $\gamma$.
Note that jump sizes are distributed normally with mean $m$ and variance $\delta^2$, and $\nu$ is finite, that is, $\nu(\bbR_0)<\infty$.
Obviously, $X$ satisfies Assumption \ref{ass1} (1) for any $\alpha>0$.
For any fixed $\alpha>0$, $\l|\int_{\bbR_0}(e^{-z_vx}-1+z_vx)\nu(dx)\r|$ is bounded on $v$, which implies
\[
|\phi(t,iz_v)|\leq C\exp\l\{-\frac{\sigma^2v^2(T-t)}{2}\r\}
\]
for some constant $C>0$.
Thus, (2) is also satisfied.
\end{ex}

\begin{ex}[Variance gamma processes]\label{ex-VG}
When $\sigma=0$ and
\[
\nu(dx)=C\l({\bf 1}_{\{x<0\}}e^{Gx}+{\bf 1}_{\{x>0\}}e^{-Mx}\r)\frac{dx}{|x|}
\]
with $C, G, M>0$, $X$ is called a variance gamma (VG) process.
For any $\alpha\in(0,M)$, $X$ satisfies Assumption \ref{ass1} (1), but (2) is not satisfied in general, since we have
\[
|\phi(t,iz_v)|\leq C_{VG}|v|^{-2C(T-t)}
\]
for some constant $C_{VG}>0$ from the view of Proposition 4.7 in Arai et al. \cite{AIS}.
\end{ex}

\begin{ex}[Normal inverse Gaussian processes]\label{ex-NIG}
$X$ is called a normal inverse Gaussian (NIG) process, if $\sigma=0$ and
\[
\nu(dx)=\frac{\delta a}{\pi}\frac{e^{bx}K_1(a|x|)}{|x|}dx,
\]
where $a>0$, $-a<b<a$, $\delta>0$, and $K_1$ is the modified Bessel function of the second kind with parameter $1$.
For more details on the function $K_1$, see Appendix A of \cite{CT}.
Since
\[
K_1(x)=e^{-x}\sqrt{\frac{\pi}{2x}}\l(1+O(x^{-1})\r)
\]
when $x\to\infty$, Assumption \ref{ass1} (1) is satisfied for $\alpha\in(0,a-b)$.
In addition, taking $\alpha\in(0,a-b)$ arbitrarily, we can find a constant $C>0$ such that
\[
\l|\int_{\bbR_0}(e^{-z_vx}-1+z_vx)\nu(dx)\r|\leq C(1+|z_v|) \ \ \ \mbox{ and } \ \ \ |\phi(t,iz_v)|\leq Ce^{-(T-t)\delta|v|}
\]
for any $v\in\bbR$ from the view of Section 5.3.8 of Schoutens \cite{Scho}.
As a result, (2) also holds.
\end{ex}

Henceforth, we fix $\alpha>0$ satisfying Assumption \ref{ass1} (1) arbitrarily.
Here we impose assumptions related to the function $f$ additionally as follows:

\begin{ass}\label{ass2}
\begin{enumerate}\renewcommand{\labelenumi}{(\arabic{enumi})}
\item $f(X_T)\in L^2(\bbP)$.
\item $f(x)e^{-\alpha x}$ is an $L^1(\bbR)$ function with finite variation on $\bbR$.
\end{enumerate}
\end{ass}

\subsection{Main theorem}
The following is a Clark-Ocone type formula for $f(X_T)$. Its proof is postponed until the next subsection.

\begin{thm}\label{thm-main}
Under Assumptions \ref{ass1} and \ref{ass2}, $f(X_T)$ is represented as
\begin{align}\label{eq-thm-main-1}
f(X_T) &= \bbE[f(X_T)]+\int_0^T\frac{\partial F}{\partial x}(s,X_s)\sigma dW_s \nonumber \\
       &  \hspace{5mm}+\int_0^T\int_{\bbR_0}\Big(F(s,X_{s-}+y)-F(s,X_{s-})\Big)\tN(ds,dy),
\end{align}
where the function $F$ is defined as
\begin{equation}\label{def-f}
F(t,x) := \bbE[f(X_T)|X_t=x] = \bbE[f(X_T-X_t+x)]
\end{equation}
for $(t,x)\in [0,T]\times\bbR$.
\end{thm}

\begin{rem}
As mentioned in Introduction, the Clark-Ocone theorem (see, e.g., Theorem 3.5.2 of \cite{Delong}) gives the same type of representation as \eqref{eq-thm-main-1}:
\begin{align*}
f(X_T) &= \bbE[f(X_T)]+\int_0^T\bbE[D_{s,0}f(X_T)|\calF_s]\sigma dW_s \\
       &  \hspace{5mm}+\int_0^T\int_{\bbR_0}\bbE[xD_{s,x}f(X_T)|\calF_{s-}]\tN(ds,dx),
\end{align*}
when $f(X_T)\in\bbD^{1,2}$.
Note that the Malliavin derivative operator $D_{s,x}$ for $(s,x)\in[0,T]\times\bbR$ and the space $\bbD^{1,2}$ are defined in Section 2.2 of \cite{Delong}.
For example, Proposition 2.6.4 of \cite{Delong} implies that $f(X_T)\in\bbD^{1,2}$ if $f$ is Lipschitz continuous and $X_T$ has a continuous density.
Thus, taking the absolute value function as $f$, we have
\begin{align*}
|X_T| &= \bbE[|X_T|]+\int_0^T\bbE[\sgn(X^\prime_{T-s}+X_s)|X_s]\sigma dW_s \\
      &  \hspace{5mm}+\int_0^T\int_{\bbR_0}\bbE\Big[|X^\prime_{T-s}+X_{s-}+y|-|X^\prime_{T-s}+X_{s-}|\Big|X_{s-}\Big]\tN(ds,dy),
\end{align*}
where $X^\prime_{T-s}$ is an independent copy of $X_T-X_s$.
This expression can be derived from not only the Clark-Ocone theorem, but also Theorem \ref{thm-main} as far as Assumption \ref{ass1} is satisfied.
Remark that we need to decompose $|X_T|$ into $X_T{\bf 1}_{\{X_T>0\}}$ and $-X_T{\bf 1}_{\{-X_T>0\}}$ in order to get the above expression via Theorem \ref{thm-main}.
On the other hand, when $f(X_T)\notin\bbD^{1,2}$, the Clark-Ocone theorem is not available,
but Theorem \ref{thm-main} is still available as far as Assumptions \ref{ass1} and \ref{ass2} are satisfied.
Some examples of such cases will be discussed in Section 2.4 below.
\end{rem}

From \eqref{eq-thm-main-5} and \eqref{eq-thm-main-7} appeared in Section 2.3 below, we can rewrite \eqref{eq-thm-main-1} as follows:

\begin{cor}\label{cor1}
Under Assumptions \ref{ass1} and \ref{ass2}, $f(X_T)$ is represented as
\begin{align*}\label{eq-cor}
f(X_T) &= \bbE[f(X_T)]+\int_0^T\frac{1}{2\pi}\int_\bbR(-z_v)\whg(X_s,-iz_v)\phi(s,iz_v)dv\sigma dW_s \nonumber \\
       &  \hspace{5mm}+\int_0^T\int_{\bbR_0}\frac{1}{2\pi}\int_\bbR\whg(X_{s-},-iz_v)\phi(s,iz_v)(e^{-z_vy}-1)dv\tN(ds,dy),
\end{align*}
where the function $\whg(x,z)$ for $(x,z)\in\bbR\times\bbC$ is defined as
\begin{equation}\label{eq-whg}
\whg(x,z):=\int_\bbR e^{izy}f(x+y)dy=e^{-izx}\whg(0,z).
\end{equation}
\end{cor}

\begin{rem}
Theorem 14.9 of Di Nunno et al. \cite{DOP} introduced the same result as Corollary \ref{cor1} for pure jump L\'evy processes, that is, the case of $\sigma=0$,
but it has not been generalized to the case of $\sigma>0$ as far as we know.
(Probably this generalization is possible by using Theorem 14.15 of \cite{DOP}.)
Note that their argument is based on the L\'evy-Wick calculus, much different from our approach.
The result of Corollary \ref{cor1} is very useful to develop a numerical scheme based on fast Fourier transform.
\end{rem}

\subsection{Proof of Theorem \ref{thm-main}}
First of all, we show $F\in C^{1,2}((0,T)\times\bbR)$.
Fix $t\in (0,T)$ and $x\in\bbR$ arbitrarily.
Remark that $F$ defined in \eqref{def-f} is represented as
\[
F(t,x)=\frac{1}{2\pi}\int_\bbR\whg(x,-iz_v)\phi(t,iz_v)dv
\]
by Proposition 2 in Tankov \cite{Tankov}, where $\whg(x,z)$ is defined in \eqref{eq-whg}.
Assumption \ref{ass2} (2) ensures that there exists a constant $\whC>0$ such that
\begin{equation}\label{eq-thm-main-2}
|z_v\whg(0,-iz_v)|<\whC,
\end{equation}
which implies that, for any $\olt\in\l[\frac{t}{2},\frac{T+t}{2}\r]$,
\begin{align*}
&\l|\whg(x,-iz_v)\frac{\partial \phi}{\partial t}(\olt,iz_v)\r| \\
&=    \l|e^{-z_vx}\whg(0,-iz_v)\phi(\olt,iz_v)(-\psi(iz_v))\r| \\
&\leq \whC e^{\alpha x}\l|\phi(\olt,iz_v)\r|\l(|\mu|+\frac{\sigma^2}{2}|z_v|+\frac{1}{|z_v|}\l|\int_{\bbR_0}(e^{-z_vy}-1+z_vy)\nu(dy)\r|\r) \\
&\leq \whC e^{\alpha x}\olh_t(v)
\end{align*}
for some integrable function $\olh_t$ by \eqref{eq-psi} and Assumption \ref{ass1} (2).
Hence, Theorem 2.27 b in Folland \cite{Folland} provides that $\displaystyle{\frac{\partial F}{\partial t}(t,x)}$ exists on $(0,T)\times\bbR$, and
\begin{align}\label{eq-thm-main-3}
\frac{\partial F}{\partial t}(t,x)
&= \frac{1}{2\pi}\int_\bbR\whg(x,-iz_v)\frac{\partial\phi}{\partial t}(t,iz_v)dv \nonumber \\
&= \frac{1}{2\pi}\int_\bbR\whg(x,-iz_v)\phi(t,iz_v)(-\psi(iz_v))dv \nonumber \\
&= \frac{1}{2\pi}\int_\bbR\whg(x,-iz_v)\phi(t,iz_v)\l(\mu z_v-\frac{\sigma^2}{2}z_v^2-\int_{\bbR_0}(e^{-z_vy}-1+z_vy)\nu(dy)\r)dv
\end{align}
holds.
Next, we focus on $\displaystyle{\frac{\partial F}{\partial x}}$ and $\displaystyle{\frac{\partial^2 F}{\partial x^2}}$.
Note that
\[
\frac{\partial\whg}{\partial x}(x,-iz_v)=-z_ve^{-z_vx}\whg(0,-iz_v)=-z_v\whg(x,-iz_v).
\]
Thus, for any $\olx\leq x$, Assumption \ref{ass1} (2), together with \eqref{eq-thm-main-2}, implies that 
\[
\l|\frac{\partial\whg}{\partial\olx}(\olx,-iz_v)\phi(t,iz_v)\r| = e^{\alpha\olx}\l|(-z_v)\whg(0,-iz_v)\phi(t,iz_v)\r| \leq \whC e^{\alpha x}\l|\phi(t,iz_v)\r|
\]
and 
\[
\l|\frac{\partial^2\whg}{\partial\olx^2}(\olx,-iz_v)\phi(t,iz_v)\r| \leq e^{\alpha x}|z_v^2\whg(0,-iz_v)\phi(t,iz_v)| \leq \whC e^{\alpha x}|z_v\phi(t,iz_v)|
\]
are integrable functions of $v$ on $\bbR$.
Therefore, we obtain that $F\in C^{1,2}((0,T)\times\bbR)$ by Theorem 2.27 in \cite{Folland}.

Secondly we show that
\begin{align}\label{eq-thm-main-4}
&\frac{\partial F}{\partial t}(t,X_t)+\frac{\partial F}{\partial x}(t,X_t)\mu+\frac{\sigma^2}{2}\frac{\partial^2 F}{\partial x^2}(t,X_t) \nonumber \\
& \hspace{5mm}+\int_{\bbR_0}\l(F(t,X_t+y)-F(t,X_t)-\frac{\partial F}{\partial x}(t,X_t)y\r)\nu(dy)=0.
\end{align}
We have
\begin{align}\label{eq-thm-main-5}
\frac{\partial F}{\partial x}(t,X_t) &= \frac{1}{2\pi}\int_\bbR\frac{\partial\whg}{\partial x}(X_t,-iz_v)\phi(t,iz_v)dv \nonumber \\
                                     &= \frac{1}{2\pi}\int_\bbR(-z_v)\whg(X_t,-iz_v)\phi(t,iz_v)dv,
\end{align}
and 
\begin{equation}\label{eq-thm-main-6}
\frac{\partial^2 F}{\partial x^2}(t,X_t) = \frac{1}{2\pi}\int_\bbR z_v^2\whg(X_t,-iz_v)\phi(t,iz_v)dv.
\end{equation}
Noting that
\[
F(t,X_t+y)=\frac{1}{2\pi}\int_\bbR\whg(X_t,-iz_v)\phi(t,iz_v)e^{-z_vy}dv
\]
holds for any $y\in\bbR$, we have 
\begin{align}\label{eq-thm-main-7}
&{F(t,X_t+y)-F(t,X_t)-\frac{\partial F}{\partial x}(t,X_t)y} \nonumber \\
& \hspace{5mm}=\frac{1}{2\pi}\int_\bbR\whg(X_t,-iz_v)\phi(t,iz_v)(e^{-z_vy}-1+z_vy)dv.
\end{align}
Hence, \eqref{eq-thm-main-4} holds from \eqref{eq-thm-main-3} and \eqref{eq-thm-main-5}--\eqref{eq-thm-main-7}.

Finally, since $F\in C^{1,2}((0,T)\times\bbR)$, It\^o's formula (see, e.g., Theorem 9.4 in \cite{DOP}) is available.
Hence, \eqref{eq-thm-main-4} implies
\begin{align*}
f(X_T) &= F(T,X_T) \\
       &= F(0,X_0)+\int_0^T\frac{\partial F}{\partial t}(s,X_s)ds+\int_0^T\frac{\partial F}{\partial x}(s,X_s)\mu ds \\
       &  \hspace{5mm}+\int_0^T\frac{\partial F}{\partial x}(s,X_s)\sigma dW_s+\frac{1}{2}\int_0^T\frac{\partial^2 F}{\partial x^2}(s,X_s)\sigma^2ds \\
       &  \hspace{5mm}+\int_0^T\int_{\bbR_0}\l(F(s,X_s+y)-F(s,X_s)-\frac{\partial F}{\partial x}(s,X_s)y\r)\nu(dy)ds \\
       &  \hspace{5mm}+\int_0^T\int_{\bbR_0}\Big(F(s,X_{s-}+y)-F(s,X_{s-})\Big)\tN(ds,dy) \\
       &= F(0,X_0)+\int_0^T\frac{\partial F}{\partial x}(s,X_s)\sigma dW_s \\
       &  \hspace{5mm}+\int_0^T\int_{\bbR_0}\Big(F(s,X_{s-}+y)-F(s,X_{s-})\Big)\tN(ds,dy),
\end{align*}
from which Theorem \ref{thm-main} follows.

\subsection{Examples}
Here we illustrate martingale representations for various examples of $f$ by using Theorem \ref{thm-main} and Corollary \ref{cor1}.

\begin{ex}[Polynomial functions of $X_T$]
When $f$ is a polynomial function, it does not have the Lipschitz continuity basically,
but we can see that $f(X_T)\in\bbD^{1,2}$ under Assumption \ref{ass1} by using Proposition 2.5 of Suzuki \cite{Suz}.
Thus, we can obtain the following representation by not only Theorem \ref{thm-main} but also the Clark-Ocone theorem:
\begin{align*}
f(X_T) &= \bbE[f(X_T)]+\int_0^T\bbE[f^\prime(X^\prime_{T-s}+X_s)|X_s]\sigma dW_s \\
       &  \hspace{5mm}+\int_0^T\int_{\bbR_0}\bbE[f(X^\prime_{T-s}+X_{s-}+y)-f(X^\prime_{T-s}+X_{s-})|X_{s-}]\tN(ds,dy),
\end{align*}
where $X^\prime_{T-s}$ is an independent copy of $X_T-X_s$.
\end{ex}

\begin{ex}[$\sqrt{|X_T|}$]\label{ex-sqrt}
We introduce a martingale representation of $\sqrt{|X_T|}$ by using Theorem \ref{thm-main} or Corollary \ref{cor1}.
Note that the Clark-Ocone theorem is not available in this case, since we cannot expect that $\sqrt{|X_T|}\in\bbD^{1,2}$ when $\sigma>0$.
Suppose that $X$ satisfies Assumption \ref{ass1}.
We have then $\bbE[|X_T|]<\infty$, which ensures Assumption \ref{ass2} (1).
Since $\sqrt{|x|}$ does not satisfy Assumption \ref{ass2} (2) for any $\alpha>0$, we decompose it into $\sqrt{x\vee0}(=:f_+(x))$ and $\sqrt{(-x)\vee0}(=:f_-(x))$.
For functions $f^+$ and $f^-$, denoting
\[
Df_\pm(x):=\l\{
       \begin{array}{rl}
       f_\pm^\prime(x), & \mbox{if }x\neq0, \\
       0,               & \mbox{if }x=0,
\end{array}\r.
\]
we have and denote
\begin{align*}
-iz\whg_\pm(x,z) &:= -iz\int_\bbR e^{izy}f_\pm(x+y)dy     = -ize^{-izx}\int_\bbR e^{izy}f_\pm(y)dy \\
                 &=  e^{-izx}\int_\bbR e^{izy}Df_\pm(y)dy = \int_\bbR e^{izy}Df_\pm(x+y)dy =: \whg_{D\pm}(x,z)
\end{align*}
for $(x,z)\in\bbR\times\bbC$.
Thus, we have
\begin{align*}
\frac{\partial}{\partial x}\bbE[f_\pm(X^\prime_{T-s}+x)]\Big|_{x=X_s}
&= \frac{1}{2\pi}\int_\bbR(-z_v)\whg_\pm(X_s,-iz_v)\phi(s,iz_v)dv \\
&= \frac{1}{2\pi}\int_\bbR\whg_{D\pm}(X_s,-iz_v)\phi(s,iz_v)dv,
\end{align*}
which implies
\begin{align*}
f_\pm(X_T) &= \bbE[f_\pm(X_T)]+\int_0^T\frac{1}{2\pi}\int_\bbR\whg_{D\pm}(X_s,-iz_v)\phi(s,iz_v)dv\sigma dW_s \\
           &  \hspace{5mm}+\int_0^T\int_{\bbR_0}\bbE[f_\pm(X^\prime_{T-s}+X_{s-}+y)-f_\pm(X^\prime_{T-s}+X_{s-})|X_{s-}]\tN(ds,dy)
\end{align*}
by Theorem \ref{thm-main} or Corollary \ref{cor1}.
Therefore, since $\sqrt{|X_T|}=f_+(X_T)+f_-(X_T)$, we have
\begin{align}\label{eq-ex-sqrt}
\sqrt{|X_T|} &= \bbE\l[\sqrt{|X_T|}\r]+\int_0^T\frac{1}{2\pi}\int_\bbR\whg_D(X_s,-iz_v)\phi(s,iz_v)dv\sigma dW_s \nonumber \\
             &  \hspace{5mm}+\int_0^T\int_{\bbR_0}\bbE\l[\sqrt{|X^\prime_{T-s}+X_{s-}+y|}-\sqrt{|X^\prime_{T-s}+X_{s-}|}\Big|X_{s-}\r]\tN(ds,dy),
\end{align}
where $\displaystyle{\whg_D(x,z):=\int_\bbR e^{izy}\Big(Df_+(x+y)+Df_-(x+y)\Big)dy}$.
Remark that we cannot rewrite the second term of the above \eqref{eq-ex-sqrt} into the conditional expectation
\[
\int_0^T\bbE[Df_+(X^\prime_{T-s}+X_s)+Df_-(X^\prime_{T-s}+X_s)|X_s]\sigma dW_s,
\]
since $Df_\pm(\pm x)e^{-\alpha x}$ are not finite variation for any $\alpha>0$.
\end{ex}

\begin{ex}[${\bf 1}_{\{X_T\geq c\}}$]\label{ex-ind}
We take an indicator function as $f$, that is, $f(x)={\bf 1}_{\{x\geq c\}}$ for $c\in\bbR$.
As seen in Section 3.4, $f(X_T)\notin\bbD^{1,2}$ when $\sigma=0$ or $\int_0^\infty|x|\nu(dx)=\infty$.
Here, we illustrate a martingale representation of ${\bf 1}_{\{X_T\geq c\}}$ by using Theorem \ref{thm-main}.
Suppose that $X$ satisfies Assumption \ref{ass1}.
On the other hand, Assumption \ref{ass2} is automatically satisfied.
Denoting
\[
F(t,x):=\bbE[{\bf 1}_{\{X_T\geq c\}}|X_t=x]=\bbE[{\bf 1}_{\{X_T-X_t\geq c-x\}}]=\bbP(X_T-X_t\geq c-x),
\]
we have
\[
\frac{\partial F}{\partial x}(t,x)=p_t(c-x)=\frac{1}{2\pi}\int_\bbR(-z_v)\whg(x,-iz_v)\phi(t,iz_v)dv,
\]
where $p_t$ is the density function of $X_T-X_t$, and
\[
\whg(x,z)=-\frac{1}{iz}e^{iz(c-x)}
\]
Note that Assumption \ref{ass1} (2) ensures the existence of $p_t$.
Theorem \ref{thm-main} implies then the following martingale representation:
\begin{align*}
{\bf 1}_{\{X_T\geq c\}} &= \bbP(X_T\geq c)+\int_0^Tp_s(c-X_s)\sigma dW_s \\
                        &  \hspace{5mm}+\int_0^T\int_{\bbR_0}\Big(\bbP(X^\prime_{T-s}\geq c-X_{s-}-y|X_{s-}) \\
                        &  \hspace{5mm}-\bbP(X^\prime_{T-s}\geq c-X_{s-}|X_{s-})\Big)\tN(ds,dy).
\end{align*}
\end{ex}

\begin{ex}[$e^{X_T}{\bf 1}_{\{X_T>0\}}$]
We consider the case where $f(x)=e^x{\bf 1}_{\{x>0\}}$.
Assume that Assumption \ref{ass1} holds for some $\alpha\geq2$.
Assumption \ref{ass2} is then automatically satisfied.
Defining $F(t,x):=\bbE[f(X_T)|X_t=x]$, we have
\[
F(t,x)=e^x\int_{-x}^\infty e^yp_t(y)dy
\]
where $p_t$ is the density function of $X_T-X_t$.
Thus, we obtain
\[
\frac{\partial F}{\partial x}(t,x)=F(t,x)+p_t(-x).
\]
As a result, Theorem \ref{thm-main} provides
\begin{align*}
&e^{X_T}{\bf 1}_{\{X_T>0\}} \\
&= \bbE\l[e^{X_T}{\bf 1}_{\{X_T>0\}}\r]+\int_0^T\l(\bbE\l[e^{X^\prime_{T-s}+X_s}{\bf 1}_{\{X^\prime_{T-s}+X_s>0\}}|X_s\r]+p_s(-X_s)\r)\sigma dW_s \\
&  \hspace{5mm}+\int_0^T\int_{\bbR_0}\bbE\l[e^{X^\prime_{T-s}+X_{s-}}\l(e^y{\bf 1}_{\{X^\prime_{T-s}+X_{s-}+y>0\}}-{\bf 1}_{\{X^\prime_{T-s}+X_{s-}>0\}}\r)\Big|X_{s-}\r]\tN(ds,dy).
\end{align*}
\end{ex}

%
%
\setcounter{equation}{0}
\section{Local risk minimization for digital options}
The main goal of this section is to provide a representation of LRM strategy of digital options for exponential L\'evy models described as \eqref{eq-exp}
by using Theorem \ref{thm-main}.
Moreover, we discuss the Malliavin differentiability of ${\bf 1}_{\{X_T\geq c\}}$ in the last part of this section.

\subsection{Preparations}
We consider a financial market with maturity $T>0$, which is composed of one risk-free asset with interest rate $r\geq0$ and one risky asset.
The risky asset price at time $t\in[0,T]$ is described as
\[
S_t:=e^{rt+X_t},
\]
where $X$ is a L\'evy process given by \eqref{eq-X}.
Moreover, we denote by $\whS$ the discounted asset price process, that is, $\whS_t:=e^{-rt}S_t$, which is also given as a solution to the following stochastic differential equation:
\begin{equation}\label{SDE}
d\whS_t=\whS_{t-}\l(\whmu dt+\sigma dW_t+\int_{\bbR_0}(e^x-1)\tN(dt,dx)\r),
\end{equation}
where
\[
\whmu:=\mu+\frac{\sigma^2}{2}+\int_{\bbR_0}(e^x-1-x)\nu(dx).
\]

Next, we give a definition of LRM strategy.
The following definition is a simplified version based on Theorem 1.6 of Schweizer \cite{Sch3},
since the original one introduced by Schweizer \cite{Sch} and \cite{Sch3} is rather complicated.
Note that \cite{Sch3} treated the problem under the assumption that $r=0$. For the case where $r>0$, see, e.g., Biagini and Cretarola \cite{BC}.

\begin{defn}
\begin{enumerate}\renewcommand{\labelenumi}{(\arabic{enumi})}
\item A strategy is defined as a pair $\vp=(\xi, \eta)$, where $\xi$ is a predictable process satisfying
      \begin{equation}\label{eq-L2}
      \bbE\l[\int_0^T\xi^2_sd\la\whS\ra_s\r]<\infty,
      \end{equation}
      and $\eta$ is an adapted process such that the discounted value of $\vp$ at time $t\in[0,T]$, defined as $\whV_t(\vp):=\xi_t\whS_t+\eta_t$
      is a right continuous process with $\bbE[\whV_t^2(\vp)]<\infty$ for every $t\in[0,T]$.
      Note that $\xi_t$ and $\eta_t$ represent the amount of units of the risky and the risk-free assets respectively which an investor holds at time $t$.
\item For a strategy $\vp$, a process $\whC(\vp)$ defined by
      \[
      \whC_t(\vp):=\whV_t(\vp)-\int_0^t\xi_sd\whS_s
      \]
      for $t\in[0,T]$ is called the discounted cost process of $\vp$.
      A strategy $\vp$ is said to be self-financing if $\whC(\vp)$ is a constant.
\item Let $H$ be a square integrable random variable representing the payoff of a contingent claim at the maturity $T$.
      A strategy $\vp$ is called locally risk-minimizing (LRM) strategy for $H$, if it replicates $H$, that is,
      it satisfies $\whV_T(\vp^H)=\whH$, and $[\whC(\vp^H),\whM]$ is a uniformly integrable martingale, where $\whM$ is the martingale part of $\whS$.
\end{enumerate}
\end{defn}

\noindent
Roughly speaking, a strategy $\vp^H=(\xi^H,\eta^H)$, which is not necessarily self-financing, is called LRM strategy for $H$,
if it is the replicating strategy minimizing a risk caused by $\whC(\vp^H)$ in the $L^2$-sense among all replicating strategies.
Proposition 5.2 of \cite{Sch3} provides that, under the so-called structure condition (SC),
an LRM strategy $\vp^H=(\xi^H,\eta^H)$ for $H\in L^2(\bbP)$ exists if and only if $\whH$($=e^{-rT}H$) admits a F\"ollmer-Schweizer decomposition, that is,
$\whH$ has the following decomposition
\begin{equation}\label{eq-FS}
\whH=\whH_0+\int_0^T\xi^{FS}_sd\whS_s+L^{FS}_T,
\end{equation}
where $\whH_0\in\bbR$, $\xi^{FS}$ is a predictable process satisfying (\ref{eq-L2}) and $L^{FS}$ is a square-integrable martingale orthogonal to $\whM$ with $L^{FS}_0=0$.
Moreover, $\vp^H$ is given by
\[
\xi^H_t=\xi^{FS}_t,\hspace{3mm}\eta^H_t=\whH_0+\int_0^t\xi^H_sd\whS_s+L^{FS}_t-\xi^H_t\whS_t.
\]
As a result, it suffices to obtain a representation of $\xi^H$ or, equivalently, $\xi^{FS}$ in order to get $\vp^H$.
Thus, we identify $\xi^H$ with $\vp^H$ in this paper.

To discuss LRM strategy, we need to consider minimal martingale measure (MMM), denoted by $\tP$.
It is defined as an equivalent martingale measure under which any square-integrable $\bbP$-martingale orthogonal to $\whM$ remains a martingale.
Thus, $L^{FS}$ appeared in \eqref{eq-FS} is characterized as a martingale not only under $\bbP$ but also under $\tP$, and orthogonal to $\whM$, that is, $\la L^{FS},\whM\ra=0$.
The density of $\tP$ is given as
\begin{align*}
\frac{d\tP}{d\bbP} &= \exp\Bigg\{-\frac{\whmu\sigma}{\sigma^2+C_2}W_T-\frac{\whmu^2\sigma^2}{2(\sigma^2+C_2)^2}T \\
                   &  \hspace{5mm}+\int_{\bbR_0}\log\l(1-\frac{\whmu(e^x-1)}{\sigma^2+C_2}\r)\tN([0,T],dx) \\
                   &  \hspace{5mm}+T\int_{\bbR_0}\l(\log\l(1-\frac{\whmu(e^x-1)}{\sigma^2+C_2}\r)+\frac{\whmu(e^x-1)}{\sigma^2+C_2}\r)\nu(dx)\Bigg\},
\end{align*}
where $C_2:=\int_{\bbR_0}(e^x-1)^2\nu(dx)$.
Note that $C_2$ is finite and $\tP$ exists under Assumption \ref{ass3} below.
Moreover, by the Girsanov theorem,
\begin{equation}\label{eq-W*}
W^*_t:=W_t+\frac{\whmu\sigma}{\sigma^2+C_2}t
\end{equation}
and
\begin{equation}\label{eq-tN*}
\tN^*([0,t],dx):=\tN([0,t],dx)+\frac{\whmu(e^x-1)}{\sigma^2+C_2}\nu(dx)t
\end{equation}
are a $\tP$-Brownian motion and the compensated Poisson random measure of $N$ under $\tP$, respectively.
We can then rewrite \eqref{SDE} as
\[
d\whS_t=\whS_{t-}\l(\sigma dW^*_t+\int_{\bbR_0}(e^x-1)\tN^*(dt,dx)\r).
\]
Remark that $X$ is a L\'evy process even under $\tP$, and the L\'evy measure under $\tP$ is given as
\[
\nu^*(dx):=\l(1-\frac{\whmu(e^x-1)}{\sigma^2+C_2}\r)\nu(dx).
\]

\subsection{Main theorem}
We shall show a representation of LRM strategy for digital options by using Theorem \ref{thm-main} under $\tP$.
Thus, we need to rewrite Assumption \ref{ass1} into one under $\tP$.
Note that, as mentioned in Example \ref{ex-ind}, Assumption \ref{ass2} is automatically satisfied.

\begin{ass}\label{ass3}
\begin{enumerate}\renewcommand{\labelenumi}{(\arabic{enumi})}
\item $\int_{\bbR_0}(e^x-1)^2\nu(dx)(=C_2)<\infty$, which implies that $\bbE_{\tP}\l[e^{\alpha X_T}\r]<\infty$ holds for some $\alpha\geq1$.
      Such an $\alpha$ is fixed throughout this section.
\item $0\geq\whmu>-\sigma^2-C_2$.
\item For any $t\in[0,T)$, there exists an integrable function $h^*_t(v)$ on $\bbR$ such that
      \[
      |\phi^*(\olt,iz_v)|\l(1+|z_v|+\frac{1}{|z_v|}\l|\int_{\bbR_0}(e^{-z_vx}-1+z_vx)\nu^*(dx)\r|\r)\leq h^*_t(v)
      \]
      for $\olt\in[\frac{t}{2},\frac{T+t}{2}]$, where $\phi^*(t,z):=\bbE_{\tP}[e^{iz(X_T-X_t)}]$ for $z\in\bbC$.
\end{enumerate}
\end{ass}

\noindent
Note that Assumption \ref{ass3} (1) ensures the structure condition (SC); and MMM $\tP$ exists as an equivalent probability measure to $\bbP$ by the above (2).
Moreover, (3) is corresponding to Assumption \ref{ass1} (2), and ensures that $X_T-X_t$ has a bounded continuous density under $\tP$, denoted by $p^*_t$.

\begin{rem}\label{rem3}
By a similar argument with Example \ref{ex-Merton}, Merton jump diffusion processes satisfy Assumption \ref{ass3} without any parameter restriction.
As for NIG processes, taking $\alpha\in(\frac{3}{2},2]$, $a>\frac{5}{2}$ and $-\frac{3}{2}<b\leq-\frac{1}{2}$,
we can see that Assumption \ref{ass3} is satisfied from the view of Arai et al. \cite{AIN}.
On the other hand, VG processes violate Assumption \ref{ass3} by a similar argument with Example \ref{ex-VG}.
For more details on this matter, see Remark \ref{rem4} below.
Note that the formulations of $\vp^*$ and $\nu^*$ are given in \cite{AIS} for Merton jump diffusion processes and VG processes, and in \cite{AIN} for NIG processes, respectively.
\end{rem}

\begin{thm}\label{thm-LRM}
Under Assumption \ref{ass3}, the LRM strategy $\xi^H$ for the digital option ${\bf 1}_{\{S_T\geq K\}}$ with $K>0$ is given by
\begin{equation}\label{eq-thm-LRM}
\xi_t^H=\frac{e^{-rT}}{\whS_{t-}(\sigma^2+C_2)}\l(\kappa_t\sigma^2+\int_{\bbR_0}\Psi^*_{t-}(K,x)(e^x-1)\nu(dx)\r)
\end{equation}
for $t\in[0,T]$.
Here
\[
\kappa_t:=p^*_t(\log K-rT-X_t)
\]
and
\[
\Psi^*_t(K,x):=\tP(X^\prime_{T-t}\geq\log K-rT-X_t-x|X_t)-\tP(X^\prime_{T-t}\geq\log K-rT-X_t|X_t),
\]
where $X^\prime_{T-t}$ is an independent copy of $X_T-X_t$.
\end{thm}

\begin{rem}\label{rem4}
By Example 3.9 of \cite{AS}, we can obtain the same result as Theorem \ref{thm-LRM} by using Malliavin calculus for L\'evy processes if ${\bf 1}_{\{S_T\geq K\}}\in\bbD^{1,2}$,
where $\bbD^{1,2}$ is defined in Section 2.2 of \cite{Delong}.
Indeed, as shown in Section 4.2 of Geiss et al. \cite{GGL}, if $\sigma=0$, $\int_{\bbR_0}|x|\nu(dx)<\infty$ and $X_t$ has a bounded density,
then we have ${\bf 1}_{\{X_T\geq c\}}\in\bbD^{1,2}$, in other words, ${\bf 1}_{\{S_T\geq K\}}\in\bbD^{1,2}$.
For example, VG processes satisfy all of these conditions, although they do not satisfy Assumption \ref{ass3} as stated in Remark \ref{rem3}.
In other words, when $X$ is a VG process, Theorem \ref{thm-LRM} is not available, but we can obtain the same result via Malliavin calculus.
The Malliavin differentiability of indicator functions will be discussed in Section 3.4 below.
\end{rem}

\subsection{Proof of Theorem \ref{thm-LRM}}
Denoting by $\olxi_t$ the right hand side of \eqref{eq-thm-LRM}, and defining a $\tP$-martingale $L^H$ with $L^H_0=0$ as
\[
L^H_t:=\bbE_{\tP}\l[e^{-rT}{\bf 1}_{\{S_T\geq K\}}-e^{-rT}\bbE_{\tP}[{\bf 1}_{\{S_T\geq K\}}]-\int_0^T\olxi_sd\whS_s\Big|\calF_t\r],
\]
we have
\begin{equation}\label{eq-thm-LRM-1}
e^{-rT}{\bf 1}_{\{S_T\geq K\}}=e^{-rT}\bbE_{\tP}\l[{\bf 1}_{\{S_T\geq K\}}\r]+\int_0^T\olxi_sd\whS_s+L^H_T.
\end{equation}
It is enough to show that \eqref{eq-thm-LRM-1} is the F\"ollmer-Schweizer decomposition of $e^{-rT}{\bf 1}_{\{S_T\geq K\}}$, the discounted value of the payoff function.
To this end, we see that $L^H$ is a $\bbP$-martingale orthogonal to $\whM$.

Defining a function $F$ on $[0,T]\times\bbR$ as
\[
F(t,x):=\bbE_{\tP}[{\bf 1}_{\{S_T\geq K\}}|X_t=x]=\tP(X_T-X_t\geq\log K-rT-x),
\]
we have
\begin{align*}
F(t,X_t) &= F(0,X_0)+\int_0^t\frac{\partial F}{\partial x}(s,X_s)\sigma dW_s^* \\
         &  \hspace{5mm}+\int_0^t\int_{\bbR_0}\Big(F(s,X_{s-}+y)-F(s,X_{s-})\Big)\tN^*(ds,dy) \\
         &= F(0,X_0)+\int_0^t\kappa_s\sigma dW_s^*+\int_0^t\int_{\bbR_0}\Psi^*_{s-}(K,y)\tN^*(ds,dy)
\end{align*}
by Assumption \ref{ass3} and Example \ref{ex-ind}.
Thus, we have
\begin{align}\label{eq-thm-LRM-3}
L^H_t &= e^{-rT}F(t,X_t)-e^{-rT}F(0,X_0)-\int_0^t\olxi_sd\whS_s \nonumber \\
      &= \int_0^te^{-rT}\kappa_s\sigma dW_s^*+\int_0^t\int_{\bbR_0}e^{-rT}\Psi^*_{s-}(K,x)\tN^*(ds,dx) \nonumber \\
      &  \hspace{5mm}-\int_0^t\olxi_s\whS_{s-}\l(\sigma dW^*_s+\int_{\bbR_0}(e^x-1)\tN^*(ds,dx)\r).
\end{align}
To show that $L^H$ is a $\bbP$-martingale, we calculate the following:
\begin{align*}
&\l(e^{-rT}\kappa_s\sigma-\olxi_s\whS_{s-}\sigma\r)\frac{\whmu\sigma}{\sigma^2+C_2} \\
&= e^{-rT}\l(\kappa_sC_2-\int_{\bbR_0}\Psi^*_{s-}(K,x)(e^x-1)\nu(dx)\r)\frac{\whmu\sigma^2}{(\sigma^2+C_2)^2}
\end{align*}
and
\begin{align*}
&\int_{\bbR_0}\l(e^{-rT}\Psi^*_{s-}(K,x)-\olxi_s\whS_{s-}(e^x-1)\r)\frac{\whmu(e^x-1)}{\sigma^2+C_2}\nu(dx) \\
&= \l(\int_{\bbR_0}e^{-rT}\Psi^*_{s-}(K,x)(e^x-1)\nu(dx)-\olxi_s\whS_{s-}C_2\r)\frac{\whmu}{\sigma^2+C_2} \\
&= e^{-rT}\l(\int_{\bbR_0}\Psi^*_{s-}(K,x)(e^x-1)\nu(dx)-\kappa_sC_2\r)\frac{\whmu\sigma^2}{(\sigma^2+C_2)^2}
\end{align*}
for $s\in[0,T]$.
Therefore, \eqref{eq-thm-LRM-3}, together with \eqref{eq-W*} and \eqref{eq-tN*}, implies that
\begin{align*}
L^H_t&= \int_0^t\l(e^{-rT}\kappa_s-\olxi_s\whS_{s-}\r)\sigma dW_s \\
     &  \hspace{5mm}+\int_0^t\int_{\bbR_0}\l(e^{-rT}\Psi^*_{s-}(K,x)-\olxi_s\whS_{s-}(e^x-1)\r)\tN(ds,dx),
\end{align*}
from which $L^H$ is a $\bbP$-martingale.

Next, we see that $L^H$ is orthogonal to $\whM$.
To this end, we have only to see $\la L^H,\whM\ra=0$.
Noting that $\whM$ is given as
\[
d\whM_t=\whS_{t-}\l(\sigma dW_t+\int_{\bbR_0}(e^x-1)\tN(dt,dx)\r),
\]
we have
\begin{align*}
&d\la L^H,\whM\ra_t \\
&= \whS_{t-}\sigma^2\l(e^{-rT}\kappa_t-\olxi_t\whS_{t-}\r)dt \\
&  \hspace{5mm}+\whS_{t-}\int_{\bbR_0}\l(e^{-rT}\Psi^*_{t-}(K,x)-\olxi_t\whS_{t-}(e^x-1)\r)(e^x-1)\nu(dx)dt \\
&= \whS_{t-}e^{-rT}\l(\kappa_t\sigma^2+\int_{\bbR_0}\Psi^*_{t-}(K,x)(e^x-1)\nu(dx)\r)dt-\olxi_t\whS_{t-}^2(\sigma^2+C_2)dt \\
&= 0.
\end{align*}

Consequently, \eqref{eq-thm-LRM-1} is the F\"ollmer-Schweizer decomposition of $e^{-rT}{\bf 1}_{\{S_T\geq K\}}$, which implies that $\xi^H=\olxi$.
This complete the proof of Theorem \ref{thm-LRM}.

\subsection{Malliavin differentiability of indicator functions}
As seen in Remark \ref{rem4}, ${\bf 1}_{\{X_T\geq c\}}\in\bbD^{1,2}$ holds true for any $c\in\bbR$ when $X$ is a VG process.
That is, we can obtain the same result as Theorem \ref{thm-LRM} for VG processes by using Example 3.9 of \cite{AS}.
On the other hand, it is known that ${\bf 1}_{\{X_T\geq c\}}\notin\bbD^{1,2}$ whenever $\sigma>0$.
In other words, if $X$ includes a Brownian component such as Merton jump diffusion processes, we need to use Theorem \ref{thm-LRM} to compute $\xi^H$ in \eqref{eq-thm-LRM}.
In addition, as seen in Proposition \ref{prop1} below, even if $\sigma=0$,
we have ${\bf 1}_{\{X_T\geq c\}}\notin\bbD^{1,2}$ as long as $\int_{\bbR_0}|x|\nu(dx)=\infty$ such as NIG processes.
As a result, we can say that Theorem \ref{thm-LRM} provides the only way to calculate LRM strategy of digital options for Merton jump diffusion and NIG processes.

\begin{prop}\label{prop1}
Let $X$ be a pure jump L\'evy process with L\'evy measure $\nu$ satisfying $\int_{[-1,1]}|x|\nu(dx)=\infty$.
In addition, suppose that $X_T$ has a bounded continuous density function $p$.
Then, we have ${\bf 1}_{\{X_T\geq c\}}\notin\bbD^{1,2}$ for all $c\in\bbR$ with $p(c)>0$.
\end{prop}

\proof
Fix $c\in\bbR$ with $p(c)>0$ arbitrarily.
Note that we can find $\ve>0$ such that $p(x)>\frac{p(c)}{2}$ for any $x\in(c-\ve,c+\ve)$.
From the view of Proposition 5.4 of Sol\'e et al. \cite{SUV}, it suffices to show that
\[
\bbE\l[\int_0^T\int_{\bbR_0}|\bfPsi_{s,x}{\bf 1}_{\{X_T\geq c\}}|^2x^2\nu(dx)ds\r]=\infty,
\]
where $\bfPsi_{s,x}$ is the increment quotient operator defined in Section 5.1 of \cite{SUV}.
Thus, we have
\begin{align*}
&\bbE\l[\int_0^T\int_{\bbR_0}|\bfPsi_{s,x}{\bf 1}_{\{X_T\geq c\}}|^2x^2\nu(dx)ds\r] \\
&=    \int_0^T\int_{\bbR_0}\bbE\l[|\bfPsi_{s,x}{\bf 1}_{\{X_T\geq c\}}|^2\r]x^2\nu(dx)ds \\
&=    \int_0^T\int_{\bbR_0}\bbE\l[\frac{|{\bf 1}_{\{X_T+x\geq c\}}-{\bf 1}_{\{X_T\geq c\}}|^2}{x^2}\r]x^2\nu(dx)ds \\
&=    \int_0^T\l(\int_0^\infty\bbP(c>X_T\geq c-x)\nu(dx)+\int_{-\infty}^0\bbP(c-x>X_T\geq c)\nu(dx)\r)ds \\
&\geq T\l(\int_0^\ve\bbP(c>X_T\geq c-x)\nu(dx)+\int_{-\ve}^0\bbP(c-x>X_T\geq c)\nu(dx)\r) \\
&\geq T\frac{p(c)}{2}\int_{(-\ve,\ve)}|x|\nu(dx)=\infty,
\end{align*}
since $\int_I|x|\nu(dx)=\infty$ for any interval $I\subset\bbR$ including $0$ as an interior point.
\fin

\begin{center}
{\bf Acknowledgments}
\end{center}
Takuji Arai gratefully acknowledges the financial support of the MEXT Grant in Aid for Scientific Research (C) No.18K03422.



\begin{thebibliography}{9999}
\bibitem{AIN}
T. Arai, Y. Imai, R. Nakashima, Numerical analysis on quadratic hedging strategies for normal inverse Gaussian models, 
in: S. Kusuoka, T. Maruyama (Eds.), Advances in Mathematical Economics, Springer, Singapore, 2018, pp. 1--24.
\bibitem{AIS}
T. Arai, Y. Imai, R. Suzuki, Numerical analysis on local risk-minimization for exponential L\'evy models, Int. J. Theor. Appl. Finance 19 (02) (2016) 1650008,
\url{https://doi.org/10.1142/S0219024916500084}.
\bibitem{AS}
T. Arai, R. Suzuki, Local risk-minimization for L\'evy markets, Int. J. Financ. Eng. 2 (02) (2015) 1550015, \url{https://doi.org/10.1142/S2424786315500152}.
\bibitem{BC}
F. Biagini, A. Cretarola, Local risk-minimization for defaultable claims with recovery process, A. Appl. Math. Optim. 65 (3) (2012) 293--314.
\url{https://doi.org/10.1007/s00245-011-9155-8}.
\bibitem{CT}
R. Cont, P. Tankov, Financial Modeling with Jump Process, Chapman \& Hall, London, 2004.
\bibitem{Delong}
\L. Delong, Backward Stochastic Differential Equations with Jumps and their Actuarial and Financial Applications, Springer, London, 2013.
\bibitem{DOP}
G. Di Nunno, B. \O ksendal and F. Proske, Malliavin Calculus for L\'evy Processes with Applications to Finance, Springer, Berlin, 2009.
\bibitem{Folland}
G.B. Folland, Real Analysis: Modern Techniques and Their Applications, 2nd ed., Wiley, Hoboken, 1999.
\bibitem{GGL}
C. Geiss, S. Geiss, E. Laukkarinen, A note on Malliavin fractional smoothness for L\'evy processes and approximation, Potential Anal. 39 (3) (2013) 203--230.
\url{https://doi.org/10.1007/s11118-012-9326-5}.
\bibitem{Sato}
K. Sato, L\'evy processes and infinitely divisible distributions, Cambridge University Press, Cambridge, 2013.
\bibitem{Scho}
W. Schoutens, L\'evy Process in Finance: Pricing Financial Derivatives, John Wiley \& Sons, Hoboken, 2003.
\bibitem{Sch}
M. Schweizer, A guided tour through quadratic hedging approaches, in: E. Jouini, J. Cvitanic, M. Musiela (Eds.), Option pricing, interest rates and risk management,
Cambridge, Cambridge University Press, 2001, pp. 538--574.
\bibitem{Sch3}
M. Schweizer, Local risk-minimization for multidimensional assets and payment streams, Banach Cent. Publ. 83 (2008) 213--229. \url{http://dx.doi.org/10.4064/bc83-0-13}.
\bibitem{SUV}
J.L. Sol\'e, F. Utzet, J. Vives, Canonical L\'evy process and Malliavin calculus, Stochastic Process. Appl. 117 (2) (2007) 165--187.
\url{https://doi.org/10.1016/j.spa.2006.06.006}.
\bibitem{Suz}
R. Suzuki, A Clark-Ocone type formula under change of measure for canonical L\'evy processes, Research Report, KSTS/RR-14/002, Keio University. \\
\url{http://www.math.keio.ac.jp/library/research/report/2014/14002.pdf}
\bibitem{Tankov}
P. Tankov, Pricing and hedging in exponential L\'evy models: review of recent results,  Paris-Princeton Lectures on Mathematical Finance 2010,
Lecture Notes in Math. 2003 (2011) 319--359. \url{https://doi.org/10.1007/978-3-642-14660-2_5}. 
\end{thebibliography}
\end{document}